\documentstyle[12pt,psfig]{article}

\begin{document}

\topmargin 0pt \oddsidemargin=-0.4truecm \evensidemargin=-0.4truecm 
\baselineskip=24pt 
\setcounter{page}{1} 
\begin{titlepage}     
\begin{flushright}
June 2001
\end{flushright}
\vspace*{0.4cm}
\begin{center}
{\LARGE\bf 
Model for Small neutrino masses at the TeV Scale} 
\end{center}
\vfil
\footnotesep = 12pt
\begin{center}
\large
Salah {\sc Nasri}$^{\it \bf a}$\footnote{
Electronic address : {\tt snasri@phy.syr.edu}} 
\quad\quad
Sherif {\sc Moussa}$^{\it \bf a,b}$\footnote{
Electronic address : {\tt sherif@phy.syr.edu}}\\
{\it 
\qquad $^{\it \bf a}$ Department of Physics, Syracuse University, 
Syracuse, NY 13244-1130, USA.} \\
\vskip 0.5cm
{ \it  \qquad $^{\it \bf b}$ 
Department of Mathematics, Faculty of Science, Ain Shams University, Cairo,11556,  
Egypt.}\\
\end{center}
\vfill
\begin{center}
\bf
\end{center}
\begin{abstract}
We propose a model for neutrino mass generation in wich no physics
beyond a TeV is required. We extend the
standard model by adding two charged singlet fields with lepton number
two.  Dirac neutrino masses $m_{\nu_D} \leq MeV$ are generated at the one loop level. Small left handed majorana neutrino
masses can be generated via the seesaw mechanism with right
handed neutrino masses  $M_R$ are of order TeV scale.

\end{abstract}
\vspace{2cm}
\centerline{} 
\vspace{.3cm}
\end{titlepage}
\renewcommand{\thefootnote}{\arabic{footnote}} \setcounter{footnote}{0} 
\newpage
 Solar and atmospheric neutrino oscillation experiments\cite{ATMSOLAR}, give a strong
 evidence for the neutrino masses and mixings. In the minimal standard
 model, neutrinos are massless, due to the abscence of the right
 handed neutrino and the conservation of lepton number. Non zero Dirac
 neutrino masses can be obtained by adding, singlet
 fermions (right handed neutrinos) to the SM. However, in this case
 one needs to fine tune the Yukawa couplings to be $\sim 10^{-12}$, which is
 quite unnatural.

An elegant and natural explanation of the smallness of the neutrino
masses compared to the quarks and charged lepton masses is provided by
the seesaw mechanism\cite{seesaw}. In this case the Majorana neutrino mass matrix
is given by:
\begin{equation}
m_\nu = -  m_D^T M_R^{-1} m_D ,
\end{equation}
where $ m_D$ is the Dirac neutrino mass matrix and $M_R$ denotes the
right handed Majorana neutrino mass matrix. This mechanism can be
implemented naturally in many extensions of the standard model gauge
group (e.g, in $SO(10)$ the right handed neutrino masses are associated
with the $(B - L)$ symmetry breaking scale)\cite{Mohapatra}.
The smallness of the neutrino mass is due to the
suppression by the scale of the heavy fields (right handed
neutrinos ) after integrating them out. An equally attractive mechanism
has been proposed recently \cite{Maetal}where the lepton number is
broken expilicitly by the coupling of a  heavy higgs triplet to the
standard model doublet.

Another interesting way to generate small Majorana neutrino masses is
given by the Zee model\cite{Zee}, in which the masses are generated at
the one loop level. In addition to the standard model higgs doublet $\Phi_{SM}$ the model
contains a charged scalar field $h^{(+)}$, that is a singlet of $SU(2)_L$,
and another doublet scalar field $\Phi$ which has the same quantum
numbers as the standard model higgs. The singlet $h^{(+)}$ carries
lepton number $-2$, so that the total lepton number L is conserved in
the Yukawa sector. The lagrangian of the Zee model reads:
\begin{equation}
{\cal L}_{Zee} = f_{\alpha \beta} L_\alpha^T Ci\tau_2 L_\beta  h^{(+)} + \mu
\Phi_{SM}^Ti\tau_2 \Phi  h^{(-)}+ h.c ,
\end{equation}
where $L_\alpha$ is the lepton doublet, C denotes the charge
conjugation, and $i\tau_2$ is the $SU(2)$ Levi-Civita symbol. Lepton
number is broken explicitly by two
unites throught the cubic coupling $\mu \Phi_{SM}^{T}i\tau_2
\Phi  h^{(-)}$. In addition, a discrete symmetry is needed such that the higgs doublet $\Phi$ does not couple to leptons. A calculable
 neutrino Majorana masses are generated at one loop through the
exchange of the physical higgs and charged lepton fields. Fermi
statistics imply that the coupling matrix $f_{\alpha \beta}$ is
antisymmetric, which leads to a neutrino mass matrix with vanishing 
diagonal elements.

The possibility of applying the radiative mechanism to generate Dirac fermion
masses, including neutrinos, have been studied previously in the
context of left right symmetric and superstring inspired models \cite{Dirac}. The authors of Ref. \cite{BM} have presented a model with
large  magnetic moment of the electron neutrino, by adding a singly
charged higgs field , in addition to the right
handed neutrinos. However, in this model the one loop neutrino Dirac mass is
logarithmically divergent and must be cancelled by adding a
counterterm, making the neutrino masses arbitrary.

In this letter we propose a model in which lepton number is broken by the right
handed Majorana mass $M_R \sim TeV$, and where calculable Dirac neutrino
masses are induced at one loop via the exchange of singlet charged
scalar fields.  

The model is the standard model extended by introducing two
singlet charged scalar fields, $S_1$ and $S_2$. No extra doublet is
needed here. We also introduce three families of right-handed neutrinos 
$\nu_{R_1},\nu_{R_2},\nu_{R_3}$. Let us assume for the moment that the total
lepton number is conserved. The most general relevant part of the
lagrangian consistent with this symmetry is given by:
\begin{equation}
{\cal L}_{ext} = f_{\alpha \beta} L_\alpha^T Ci\tau_2 L_\beta  S_1^{(+)} + 
g_{\alpha \beta}l_{R_{\alpha}}^T C \nu_{R_{\beta}}  S_2^{(+)} + h.c
\end{equation}
where, $\alpha$, $\beta$ denote generation indices. The couplings
$f_{\alpha \beta}$ are antisymmetric in $\alpha$ and $\beta$, while 
$g_{\alpha \beta}$ are arbitrary.
Conservation of lepton number requires that $S_1^{(+)}$ and
$S_2^{(+)}$ carry lepton number $L = -2$. However at this stage a coupling between the
standard model higgs and the right handed neutrino is allowed. The
presence of such term in the lagrangian leads to a Dirac mass for the
neutrino, and one needs to fine tune the Yukawa coupling to produce
$m_\nu \ll m_e$. To forbid these couplings we impose a $Z_2$ symmetry
acting as  follows: 
\begin{equation}
(L_\alpha , \Phi_{SM} , S_1^{(+)}) \longrightarrow  (L_\alpha ,
  \Phi_{SM},   S_1^{(+)})
\end{equation}
\begin{equation}
(N_{R_\alpha} , S_2^{(+)}) \longrightarrow - (N_{R_\alpha} , S_2^{(+)})
\end{equation}
In the limit where $Z_2$ is unbroken the Dirac mass for the neutrino vanishes
to any order of perturbation theory. We will therfore assume that $Z_2$ is
broken softly in the higgs sector.

 The scalar potential is defined by:
\begin{equation}
V = V_1(\Phi_{SM}\Phi_{SM}^{\dagger}, S_1S_1^{\dagger},
S_2S_2^{\dagger}) + \kappa S_1S_2^{\dagger}
\end{equation}
where $V_1(\Phi_{SM}\Phi_{SM}^{\dagger}, S_1S_1^{\dagger})$ contains the
usual terms  with positive mass square and quadratic terms for
$S_1$ and $S_2$ such that $<S_1> = <S_2> = 0$ (since they carry
electric charge). The term $\kappa S_1S_2^{\dagger}$ is the $Z_2$ soft
breaking term which can be of order of the charged singlet higgs
masses $m_S$. The physical  charged higgs fields are given in the basis $S_1^{(+)}$ and
$S_2^{(+)}$ by:
\begin{eqnarray}
h_1 = S_1^{(+)}\cos\theta + S_2^{(+)}\sin\theta & \nonumber\\
h_2 = -S_1^{(+)}\sin\theta + S_2^{(+)}\cos\theta &&
\end{eqnarray}
where 
\begin{equation}
\sin^2\ 2\theta = \frac{4\kappa^2}{4\kappa^2 + (m_2 - m_1)^2}
\end{equation}
Dirac masses of the neutrino are generated from the one loop diagrams shown
in Fig.$1$ and in the basis in which the charged lepton mass
matrix is diagonal are given by
\begin{equation}
m_{\nu_{\alpha\beta}} = g_{\alpha\gamma} m_{\gamma} f_{\gamma\beta} \sin\theta \cos\theta I(M_1, M_2, m_{\beta})
\end{equation}
where $m_{\beta}$ are the masses of the charged leptons,  $M_1$,
$M_2$ are the masses of the physical charged higgs $h_1$ and $h_2$ and
$I(M_1, M_2, m_{\beta})$ is the one loop integral given by:
\begin{equation}
I(M_1, M_2, m_{\beta}) = \int \frac{d^4 k}{(2\pi)^4} \frac{1}{k^2 - m_{\beta}^2}
\left(\frac{1}{k^2 - M_2^2} - \frac{1}{k^2 - M_1^2}\right),
\end{equation}
in evaluating this integral the charged lepton masses can be safely
 neglected so that the one loop induced Dirac neutrino masses are given by
\begin{equation}
m_{\nu_{\alpha\beta}} = \frac{1}{64\pi^2} g_{\alpha\gamma} m_{\gamma}
 f_{\gamma\beta}\sin(2\theta)\ln(\frac{M_2}{M_1})^2
\end{equation} 
In this model neutrino masses are proportional to the charged lepton
masses, while in the Zee model the Majorana neutrinos are quadratic in
the charged lepton masses. Note that one of eigenvalues of the
neutrino mass matrix in Eq. $(11)$ vanishes, since the $3\times 3$ antisymmetric matrix
$f_{\alpha\beta}$ has zero determinant. The same feature has been
noticed in another model \cite{Bab} with a doubly charged scalar. The higher loop
correction can always be written in the form $g^TYf$, where Y depends
on the parameters of the model. Thus at any higher loop level one of
the neutrinos remains massless. Note that although the charged lepton
masses receive
corrections at one loop by replacing the charged lepton fields in
Fig.$1$  by the Dirac neutrino fields, this correction is $\delta m_l
\simeq \frac{m_\nu^2}{m_l}$, which can safely be neglected.

Now let us study the phenomenological constraints on the parameters of
the model. The physical charged scalar fields $h_1$ and $h_2$ can
mediate the decays $\mu\longrightarrow e\nu_e\nu_{\mu}$,
$\tau\longrightarrow e\nu_e\nu_{\tau}$, and $\tau\longrightarrow
\mu\nu_{\mu}\nu_{\tau}$. The muon decay (see Fig. $2$) can be described by the
effective lagrangian:
\begin{equation}
{\cal L}_{eff} = -\frac{4G_F}{\sqrt2}[(1 + \delta_{L_{e\mu}})\overline
e_{L}\gamma^{\rho}\nu_{e_L}\overline\nu_{\mu}\gamma^{\rho}\mu_L +  
\delta_{RR_{e\mu}}\overline
e_{R}\gamma^{\rho}\nu_{e_R}\overline\nu_{\mu_R}\gamma^{\rho}\mu_R +
\delta_{LR_{e\mu}}\overline e_{R}\nu_{e_L}\overline\nu_{\mu_R}\mu_L ] , 
\end{equation}
where 
\begin{eqnarray}
\delta_{LL_{e\mu}} = \frac{f_{e\mu}^2}{\sqrt2G_F\overline M^2} , & \\
\delta_{RR_{e\mu}} = \frac{g_{e\mu}^2}{\sqrt2G_F\overline M^2} , & \\
\delta_{LR_{e\mu}} = \frac{f_{e\mu}g_{e\mu}}{\sqrt2G_F\overline M^2} , &&
\end{eqnarray}
with:
\begin{equation}
\overline M^2 = \frac{M_1^2M_2^2}{M_1^2\cos^2\theta + M_2^2\sin^2\theta}
\end{equation}
For the $\tau\longrightarrow e\nu_e\nu_{\tau}$ and
$\tau\longrightarrow \mu\nu_{\mu}\nu_{\tau}$ decays one has to substitute
$\delta_{e\mu}$ by $\delta_{e\tau}$ and $\delta_{\mu\tau}$
respectively. The more stringent constraint  on $f_{e\mu}$ is derived
from universality of the $\mu$ decay which gives\cite{constraints}
\begin{equation}
\frac{f_{e\mu}}{M}\leq 10^{-4}GeV^{-1}.
\end{equation} 
A useful constraint on $g_{e\mu}$ can be obtained from $e - \mu$
 universality which gives the bound
\begin{equation}
\frac{g_{e\mu} f_{e\mu}}{\overline M^2 } \leq 3\times 10^{-6}GeV^{-2}.
\end{equation}
The charged scalars $h_1$ and $h_2$ contribute to the
$\mu\longrightarrow e\gamma$ . A straightforward calculation gives:
\begin{equation}
\Gamma(\mu\longrightarrow e\gamma) \simeq
\alpha\frac{1}{384\pi^4}\frac{m_\mu^5}{\overline M^4}\mid(ff^+)_{12} +
(gg^+)_{12}\mid^2
\end{equation}
and by using the experimental upper limit $B(\mu\longrightarrow
e\gamma) < 1.2\times 10^{-11}$ we obtain
\begin{equation}
\frac {1}{\overline M^2}\mid(ff^+)_{12} +
(gg^+)_{12}\mid < 2.8\times 10^{-9}GeV^{-2}.
\end{equation}
 Another bound on the $(V +A)$ couplings $g_{\alpha\beta}$ can be
drived from the constraint on the Michel parameter $\zeta$. Using the
experimental bounds from Ref. \cite{particledata} , we obtain
\begin{equation}
\frac{g_{e\mu}^2}{\overline M^2} < 3\times 10^{-4} GeV^{-2}
\end{equation}
Note that the decay
$\mu\longrightarrow \overline\nu_e \nu_{\mu_R} e_R$ will be forbidden
kinimatically in the seesaw case, since the mass of the right handed
neutrino is much bigger than the mass of the muon. Note that for
$M_{1,2}\simeq 10$TeV and $f<1$, $g<1$ the above constraints are satisfied.

Now let us turn to the mixings and the masses relevant for neutrino
 oscillations, we assume for simplicity  $g_{\alpha\beta}\simeq
f_{\alpha\beta}$, such that the model has four remaining parameters. In this case the neutrino mass matrix is symmetric,
and has the following form:
\begin{equation} 
m_{\nu} \simeq  m_0\left[\begin{array}{c c c}
f_{e\tau}^2 + f_{e\mu}^2(\frac{m_{\mu}}{m_{\tau}})&f_{e\tau}f_{\mu\tau}&-f_{e\mu}f_{\mu\tau}(\frac{m_{\mu}}{m_{\tau}})\\
f_{e\tau}f_{\mu\tau}&f_{\mu\tau}^2&f_{e\mu}f_{e\tau}(\frac{m_e}{m_{\tau}})\\
-f_{e\mu}f_{\mu\tau}(\frac{m_{\mu}}{m_{\tau}})&f_{e\mu}f_{e\tau}(\frac{m_e}{m_{\tau}})&f_{\mu\tau}^2(\frac{m_{\mu}}{m_{\tau}}) 
\end{array}
\right]
\end{equation}
where $m_0$ corresponds to the scale of neutrino mass in this model, and
it is given by:
\begin{equation}
m_0 = \frac{1}{64\pi^2}m_{\tau}\frac{4\kappa^2}{M_2^2 + 4\kappa^2} ,
\end{equation}
where in deriving the above relation we have assumed that the charged higgs $h_{2}$ is 
heavier than $h_1$. For $\kappa\simeq M_2$, $m_0\simeq 2 MeV$, which
implies that the parameters $f_{\alpha\beta}\leq 10^{-3}$. It is well
known that a stable $MeV$ neutrino mass can be problematic for cosmology
\cite{cosmology}. Morever primordial
 nucleosynthesis excludes a Dirac neutrino mass from $.3$ to $ 25 MeV$\cite{BBN}, which implies that
$f_{\alpha\beta}$ should be less than one.  

If there is no strong hierarchy between the Yukawa couplings
 $f_{\alpha\beta}$, then the matrix element $m_{\nu_{23}}$ can be safely
 neglected, and the eigenvalues of $m_{\nu}$ are
\begin{eqnarray}
m_{\nu_1} & = & 0 \nonumber \\                                                   
m_{\nu_2} & \simeq & m_0\left(f_{\mu\tau}^2 +
\frac{1}{2}(\frac{m_{\mu}}{m_{\tau}})f_{e\mu}^2 - \sqrt{f_{e\tau}^4 + \frac{1}{4}(\frac{m_{\mu}}{m_{\tau}})^2f_{e\mu}^4}\right)   \nonumber \\
m_{\nu_3} & \simeq & m_0\left(f_{\mu\tau}^2 +
\frac{1}{2}(\frac{m_{\mu}}{m_{\tau}})f_{e\mu}^2 + \sqrt{f_{e\tau}^4 +
\frac{1}{4}(\frac{m_{\mu}}{m_{\tau}})^2 f_{e\mu}^4}\right)     
\end{eqnarray}
This means that $m_{\nu_3}$ is the scale relevant for atmospheric
neutrino oscillations, which implies that $f_{\mu\tau} \sim  f_{e\tau} \sim 10^{-4}$. However the explanation of the solar neutrino
anomaly requires a  fine tuning between $f_{e\tau}$ and  $f_{\mu\tau}$ 
such that
$m_{\nu_2}<<\Delta m_{atm}^2$. The matrix $m_{\nu}$ can be diagonalized
(neglecting CP violation in the leptonic sector) by the following orthogonal transformation:
\begin{equation}
V = \left[\begin{array}{c c c}
\frac{1}{2aC_1}&-\frac{ a + b + \sqrt{a^2 + b^2 }}{c_2}&-\frac{a + b -
\sqrt{a^2 + b^2 }}{c_3}\\
-\frac{1}{2aC_1}&\frac{a - b - \sqrt{a^2 + b^2}}{c_2}&
\frac{a - b + \sqrt{a^2 + b^2}}{c_3}\\
\frac{1}{C_1}&\frac{1}{C_2}&\frac{1}{C_3}
\end{array}
\right]
\end{equation}
where :
\begin{eqnarray}
a & = & \frac{f_{e\mu}}{2f_{\mu\tau}} \\
b & = & \frac{f_{\mu\tau}}{\epsilon f_{e\mu}}  \\
c_1 & = & \frac{\sqrt{1 + 4a^2}}{a}  \\
c_2 & \simeq & 2(a^2 + b^2 + a\sqrt{a^2 + b^2} +\frac{1}{4})^{\frac{1}{2}}  \\
c_3 & \simeq & 2(a^2 + b^2 + b\sqrt{a^2 + b^2} + \frac{1}{4})^{\frac{1}{2}} 
\end{eqnarray}
From the above mixing matrix one can see that the large mixing angle
required for the atmospheric neutrino oscillations can not be satisfied 
simultanously with the required smallness of the $U_{e3}$ element.
One could ask if the model can accomodate both the
atmospheric and solar neutrino anomalies for $g_{\alpha\beta}$ $\ne
f_{\alpha\beta}$. However we have performed an analysis of all possible parameters
$f_{\alpha\beta}$, $g_{\alpha\beta}$, and we found no solution
that can fit both atmospheric and solar neutrino puzzles with
$U_{e3}\leq 0.2$.

Now let us assume that lepton number is broken by the right handed
neutrino mass term:
\begin{equation}
{\cal L}_{\Delta L=2} = \frac{1}{2}M_{R_{\alpha\beta}}N_{R_{\alpha\beta}}^TCN_{R_{\alpha\beta}}
+ h.c
\end{equation}
The masses of the light neutrinos are obtained by diagonalizing the
following mass matrix in the basis $\nu_L, N_R^c$:
\begin{equation}
\left[\begin{array}{c c c}
0&m^T_{\nu_D}\\
m_{\nu_D}&M_R,
\end{array}
\right]
\end{equation} 
where $m_{\nu_D}$ is the Dirac neutrino mass matrix induced at one
loop as in Eq.($11$). The seesaw approximation ($M_R >> m_{\nu_D}$) implies 
\begin{equation}
M_{\nu}^{(light)} \simeq - m_{\nu_D}^TM_R^{-1}m_{\nu_D}
\end{equation}
We will assume that the Yukawa coupling constants $f_{\alpha\beta}$,
$g_{\alpha\beta}$ are of order unity, and $M_{R}$,$ M_1$ and $M_2$ in
the TeV range. In this case the light Majorana neutrino masses are smaller than
eV. Of course, one of the eigenvalues will vanishe, and this requires the
other masses to be in the atmospheric and solar neutrino ranges. This
can be easily accomodated since the seesaw fomula is very sensitive
to any deviation of the coupling constants from unity, since it varies
as the fourth power of the coupling constants.

The model can not predict the mixing angles since the parameters
$f_{\alpha\beta}$ and $g_{\alpha\beta}$ are arbitrary, in contrast to
the Zee model, in which the
neutrino mass matrix has vanishing diagonal elements without any fine
tuning or imposing  extra symmetries. However, unlike the Dirac case,
the seesaw neutrino mass matrix can accomodate the atmospheric and the
solar anomalies. Choosing the Yukawa parameters $f_{\alpha\beta}, 
g_{\alpha\beta}$ to take the values:
\begin{eqnarray}
g_{11} = g_{22} = g_{33}= g_{21}=g_{13} = g_{31} =1\nonumber\\
 g_{23} =g_{32} =0 \nonumber \\
g_{12}  \simeq -.23 \nonumber\\
f_{23}= 1\\
f_{12} = - f_{13} \simeq .5 
\end{eqnarray}
with the right handed neutrino masses $(M_{R_1}, M_{R_2},  M_{R_3}) 
\simeq (15.2, 3.3, 10)TeV$,
leads to a bimaximal mixing with large angle MSW \cite{LMSW} effect as the solution to
the solar neutrino puzzle. The numbers above suggest that one can
easily choose a texture for the matrix $g_{\alpha\beta}$, that can fit both the
atmospheric and solar neutrino data.

In summary we have proposed a model for generating small majorana
neutrino masses at the TeV scale using the seesaw mechanism. The smallness
of the neutrino masses in this model is not due to the largness of the right
handed neutrino mass, but to the smallness of the Dirac masses. The
model may also have an intersting implication for cosmology\cite{progress}.     

\section {Aknowledgment}
One of the authors (S. N) would like to thank K. S. Babu for useful
discussion, and very thanksful to  J. Schechter and M. Trodden
for careful reading of the maniscript. S. M is very thanksful to the
Egyptian Cultural Bureau for the support.

\newpage

\begin{figure}[ht]
\centerline{\psfig{file=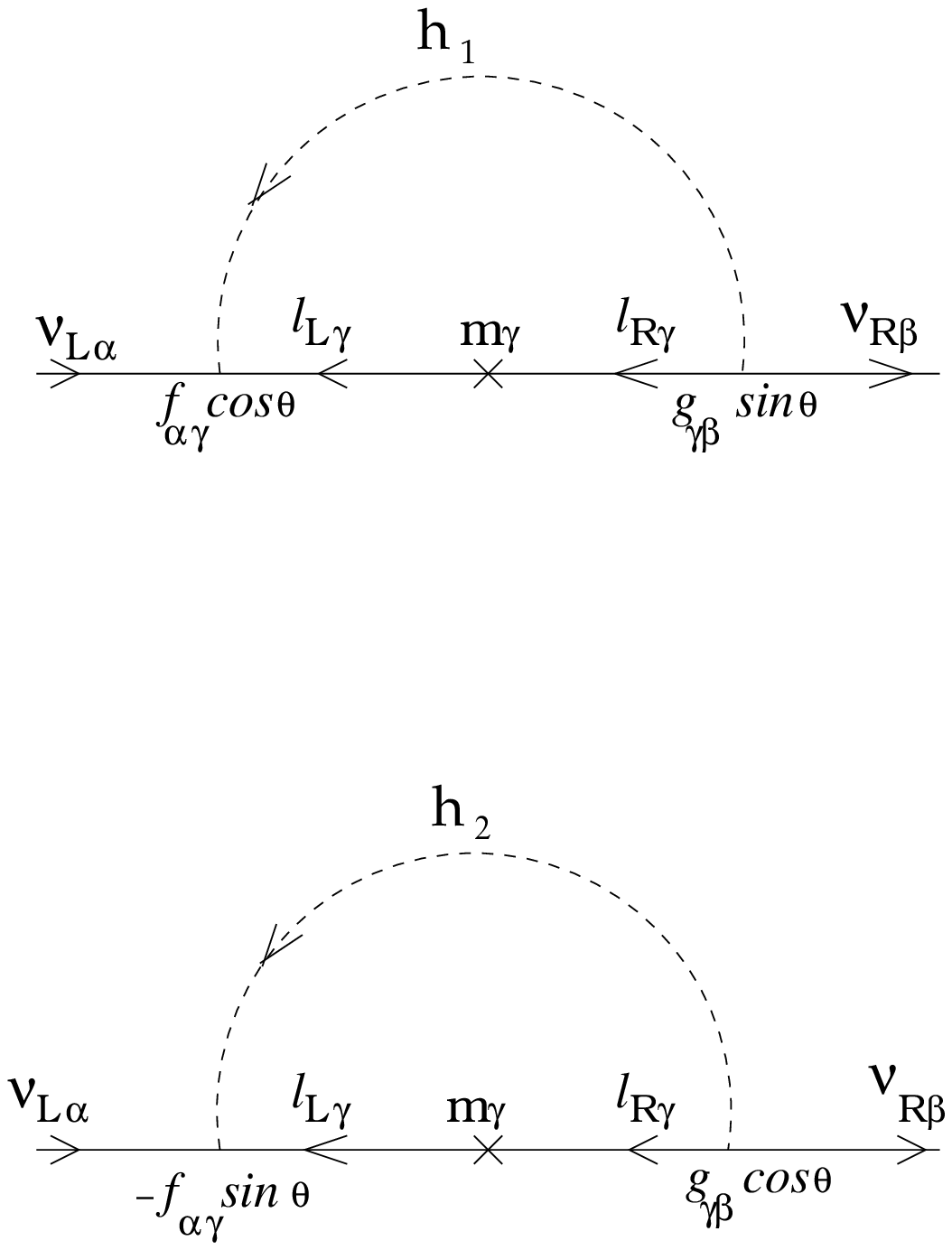,width=5in}}
\caption{One loop diagram giving rise to Dirac neutrino masses}
\end{figure}

\begin{figure}[ht]
\centerline{\psfig{file=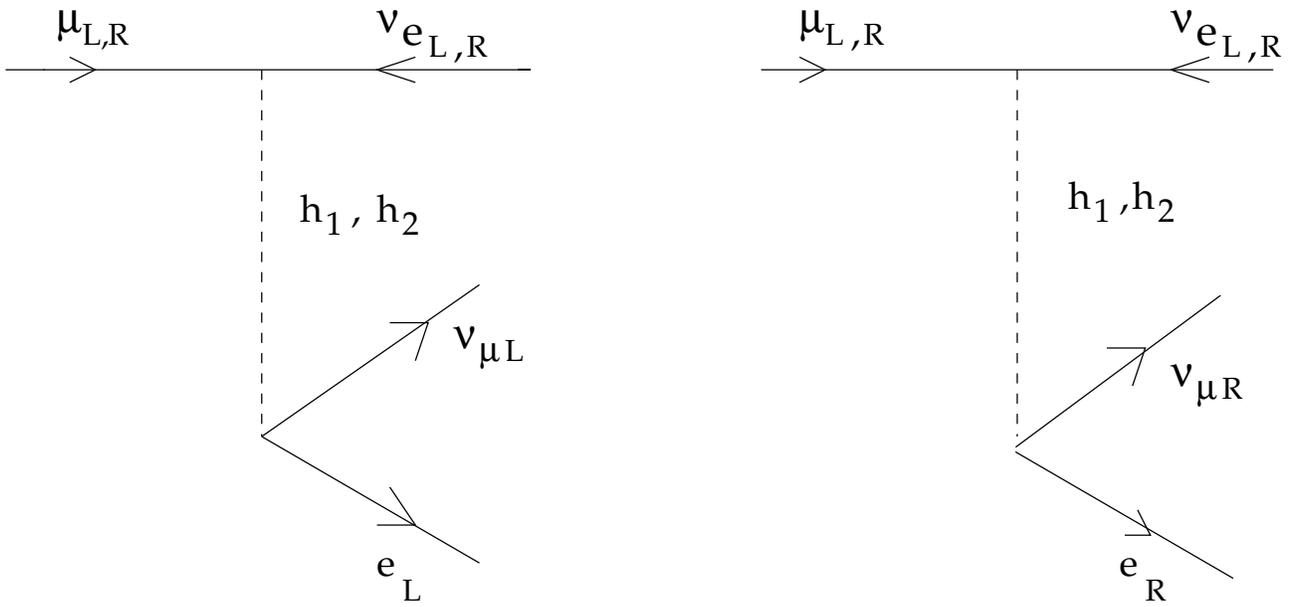,width=7in}}
\caption{Muon decay via the exchange of the charged singlet fields}
\label{}
\end{figure}

\end{document}